\documentclass[twocolumn,showpacs,prl,10pt,aps,floatfix]{revtex4-1}
\usepackage[english]{babel}
\usepackage[colorlinks,linkcolor=blue,citecolor=blue,urlcolor=blue]{hyperref}
\usepackage{amsmath}
\usepackage{graphicx}
\begin{document}
\title{Novel Magnetic Block States in Low-Dimensional Iron-Based Superconductors}
\author{J. Herbrych$^{1,2,3}$}
\author{J. Heverhagen$^{4,5}$}
\author{N. D. Patel$^{6}$}
\author{G. Alvarez$^{7}$}
\author{M. Daghofer$^{4,5}$}
\author{A. Moreo$^{1,2}$}
\author{E. Dagotto$^{1,2}$}
\affiliation{$^{1}$ Department of Physics and Astronomy, University of Tennessee, Knoxville, Tennessee 37996, USA}
\affiliation{$^{2}$ Materials Science and Technology Division, Oak Ridge National Laboratory, Oak Ridge, Tennessee 37831, USA}
\affiliation{$^{3}$ Department of Theoretical Physics, Faculty of Fundamental Problems of Technology, Wroc{\l}aw University of Science and Technology, 50-370 Wroc{\l}aw, Poland}
\affiliation{$^{4}$ Institute for Functional Matter and Quantum Technologies, University of Stuttgart, Pfaffenwaldring 57, D-70550 Stuttgart, Germany}
\affiliation{$^{5}$ Center for Integrated Quantum Science and Technology, University of Stuttgart, Pfaffenwaldring 57, D-70550 Stuttgart, Germany}
\affiliation{$^{6}$ Department of Physics, Ohio State University, Columbus, Ohio 43210, USA}
\affiliation{$^{7}$ Computational Sciences and Engineering Division and Center for Nanophase Materials Sciences, Oak Ridge National Laboratory, Oak Ridge, Tennessee 37831, USA}
\date{\today}
\begin{abstract}
Inelastic neutron scattering recently confirmed the theoretical prediction of a \mbox{$\uparrow\uparrow\downarrow\downarrow$}-magnetic state along the legs of quasi-one-dimensional (quasi-1D) iron-based ladders in the orbital-selective Mott phase (OSMP). We show here that electron-doping of the OSMP induces a whole class of novel block-states with a variety of periodicities beyond the previously reported $\pi/2$ pattern. We discuss the magnetic phase diagram of the OSMP regime that could be tested by neutrons once appropriate quasi-1D quantum materials with the appropriate dopings are identified.
\end{abstract}
\maketitle

\begin{figure}[!htb]
\includegraphics[width=1.0\columnwidth]{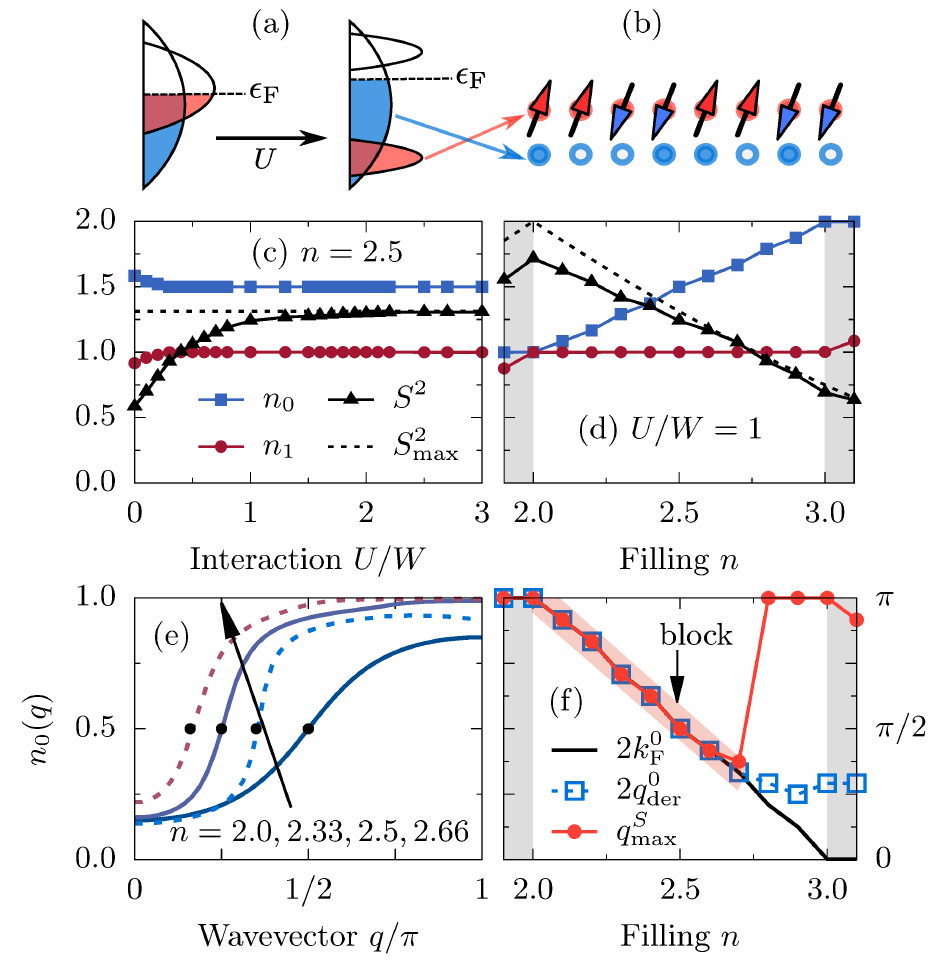}
\caption{(Color online) Schematic representation of (a) the orbital-selective Mott phase and (b) $\pi/2$-block spin state. (c) Interaction $U$ and (d) filling $n$ dependence of the total magnetic moment $\langle S^2_\ell \rangle$ and orbital-resolved occupation numbers $n_{\gamma}$ of the two-orbital Hubbard model Eq.~\eqref{hamhub}, using $L=48$ sites. In (d), the black dashed line represents the largest possible magnetic moment $S^2_{\mathrm{max}}$. (e) Momentum distribution function $n_0(q)$ of the $\gamma=0$ orbital. Points indicate $q^0_{\mathrm{der}}$, the position of the maximum in \mbox{$\mathrm{d} n_{0}(q)/\mathrm{d} q$}. (f) $n$ dependence of the position of the maximum of the spin structure factor $q^S_{\mathrm{max}}$. Block-OSMP is denoted as a colored area. In the same panel we also present the position of the orbital $\gamma=0$ Fermi-vector calculated via \mbox{$2k^0_{\mathrm{F}}=\pi n_0$} and via $2q^0_{\mathrm{der}}$.}
\label{ocu_nh}
\end{figure}

{\it Introduction.} Competing interactions in strongly correlated electronic systems can induce novel and exotic effects. For example, in the iron-based superconductors \cite{Stewart2011,Dagotto2013,Dai2015,Fernandes2017} charge, spin, and orbital degrees of freedom combine phenomena known from cuprates with those found in manganites. Prominent among these novel effects is the orbital-selective Mott phase (OSMP) \cite{Georges2013}, where interactions acting on a multi-orbital Fermi surface cause the selective localization of electrons on one of the orbitals. As a consequence, the system is in a mixed state with coexisting metallic and Mott-insulating bands [Fig.~\ref{ocu_nh}(a)].  Since the theoretical studies of the OSMP require the treatment of challenging multi-orbital models most of the investigations thus far were performed  with approximations such as the the slave-particle mean field method \cite{Biermann2005,Medici2009,Yu2013,Yu2017} or dynamical mean-field theory \cite{Koga2004,Jakobi2013,Roekeghem2016}. Here, we present unambiguous numerical evidence of the OSMP within low-dimensional multi-orbital Hubbard models, unveiling a variety of new phases.

The magnetic ordering associated with the OSMP could be significantly different from that observed in cuprates. The latter are described by single-band Hamiltonians and the parent compounds order in a staggered antiferromagnetic fashion. However, recent inelastic neutron scattering (INS) experiments \cite{Mourigal2015} on quasi-1D iron-based materials of the 123 family (AFe$_2$X$_3$; A alkali metals, X=Se,S chalcogenides) unveiled exotic \mbox{$\pi/2$-block} magnetic states where spins form antiferromagnetically (AFM) coupled ferromagnetic (FM) islands, in a $\uparrow\uparrow\downarrow\downarrow$-pattern [Fig.~\ref{ocu_nh}(b)]. Similar patterns were also reported in two dimensions with iron vacancies, such as in Rb$_{0.89}$Fe$_{1.58}$Se$_2$ \cite{wang2011} and K$_{0.8}$Fe$_{1.6}$Se$_2$ \cite{Bao2011,You2011,Yu2011,Yu2013}. For the aforementioned compounds the OSMP state is believed to be relevant \cite{Caron2012,Yu2013,Dong2014,Mourigal2015,Pizarro2018}.

Recent theoretical investigations \cite{Rincon2014,Herbrych2018} showed that a multi-orbital Hubbard model in the OSMP state properly describes the INS spin spectra of $\pi/2$-block state of powder BaFe$_2$Se$_3$ \cite{Mourigal2015}. However, the origin of block magnetism and its relation to the OSMP is an intriguing and generic question that has been barely explored. We show that the block-OSMP magnetism can develop in various shapes and sizes depending on the electron-doping, far beyond the previously reported \mbox{$\pi/2$-pattern}. Moreover, here we develop effective Hamiltonians for the OSMP which allows for an intuitive understanding of the origin of block magnetism. Our simplified, yet accurate, model allows for reliable numerical investigations of the OSMP and predicts the behavior of the maximum of the spin structure factor in electron-doped OMSP.

{\it Model.} Our conclusions are based on extensive simulations of two- and three-orbital models in chain and ladder geometries for a variety of model parameters. For clarity, consider first the two-orbital 1D Hubbard model,
\begin{eqnarray}
H_{\mathrm{H}}&=&-\sum_{\gamma,\gamma^\prime,\ell,\sigma}
t_{\gamma\gamma^\prime}
\left(c^{\dagger}_{\gamma,\ell,\sigma}c^{\phantom{\dagger}}_{\gamma^\prime,\ell+1,\sigma}+\mathrm{H.c.}\right)+
\Delta\sum_{\ell}n_{1,\ell}\nonumber\\
&+&U\sum_{\gamma,\ell}n_{\gamma,\ell,\uparrow}n_{\gamma,\ell,\downarrow}
+\left(U-5J_{\mathrm{H}}/2\right)\sum_{\ell}n_{0,\ell}n_{1,\ell}\nonumber\\
&-&2J_{\mathrm{H}}\sum_{\ell}\mathbf{S}_{0,\ell} \cdot \mathbf{S}_{1,\ell}
+J_{\mathrm{H}}\sum_{\ell}\left(P^{\dagger}_{0,\ell}P^{\phantom{\dagger}}_{1,\ell}
+\mathrm{H.c.}\right)\,,
\label{hamhub}
\end{eqnarray}
where $c^{\dagger}_{\gamma,\ell,\sigma}$ creates an electron with spin $\sigma=\{\uparrow,\downarrow\}$ at orbital $\gamma=\{0,1\}$ and site $\ell=\{1,\dots,L\}$. $t_{\gamma\gamma^\prime}$ denotes a diagonal hopping amplitude matrix in orbital space $\gamma$, with $t_{00}=-0.5$ and $t_{11}=-0.15$ in eV units. The crystal-field splitting is $\Delta=0.8$~eV (kinetic-energy band-width is $W=2.1$~eV). The local $(\gamma,\ell)$ orbital-resolved particle density is $n_{\gamma,\ell}=\sum_{\sigma}n_{\gamma,\ell,\sigma}$, $\mathbf{S}_{\gamma,\ell}$ is the local spin, and $P^{\phantom{+}}_{\gamma,\ell}=c_{\gamma,\ell,\uparrow}c_{\gamma,\ell,\downarrow}$ is the pair-hopping operator. The global filling is $n=N/L$, where $N$ is the number of electrons and $L$ the system size. $U$ is the repulsive Hubbard interaction, while $J_{\mathrm{H}}$ is the Hund exchange, fixed here to $J_{\mathrm{H}}=U/4$ \cite{Rincon2014,Rincon2014-2,Dai2012,Yin2010,Li2016,Kaushal2017,Herbrych2018,Patel2018}. All results are obtained with the density matrix renormalization group method \cite{White1992,White1993,Schollwock2005,Alvarez2009,Alvarez2018} with truncation errors smaller than $\sim10^{-7}$. Open boundary conditions were assumed.

Although the Hamiltonian Eq.~\eqref{hamhub} appears complex, it represents a generic SU(2) symmetric multiorbital system. As long as we are in the block-OSMP state, our results are {\it not} sensitive to details of the parameter sets: for example, in the Supplementary Material \cite{supp} we show that the same conclusions are drawn from a system with or without interorbital hybridization and with two or three orbitals. Ladder geometries also lead to similar findings. We also observe block states in the effective Kondo-Heisenberg model. Our results are thus generic and intrinsic of multi-orbital systems in the block-OSMP.

{\it Orbital-selective Mott phase.} Figures~\ref{ocu_nh}(c-d) present the orbital-resolved occupation numbers $n_{\gamma}$ and the total magnetic moment per site squared $\langle S_{\ell}^2 \rangle = S_{\ell}(S_{\ell}+1)$. As expected in the OSMP, increasing the interaction $U$ ``locks'' the occupation number of one of the orbitals to $n_1=1$. This behavior is observed in the $2<n<3$ region, yielding a remaining orbital $\gamma=0$ only fractionally occupied. Simultaneously, $\langle S_{\ell}^2 \rangle$ is almost fully saturated to its maximal value (given by the total filling) when $n_1=1$. Charge fluctuations \mbox{$\delta n^2_\gamma=\langle n^2_\gamma\rangle-\langle n_\gamma\rangle^2$} \cite{supp,Patel2018} indicate that the number of doublons is highly suppressed on the single-occupied orbital, different from the fractionally occupied orbital $\gamma=0$ where $\delta n^2_0\ne0$ suggests metallic character. The results above, and the orbital-resolved density-of-states (DOS) analysis~\cite{supp}, are consistent with OSMP physics in a wide range $2<n<3$. As in previous investigations \cite{Patel2018}, our block-OSMP system is in an overall metallic state for all considered fillings, albeit with a highly reduced Fermi-level DOS. It is currently unknown if the latter approximates a pseudogap or rather a weak insulator, and more detailed analysis is needed.

\begin{figure}[!htb]
\includegraphics[width=1.0\columnwidth]{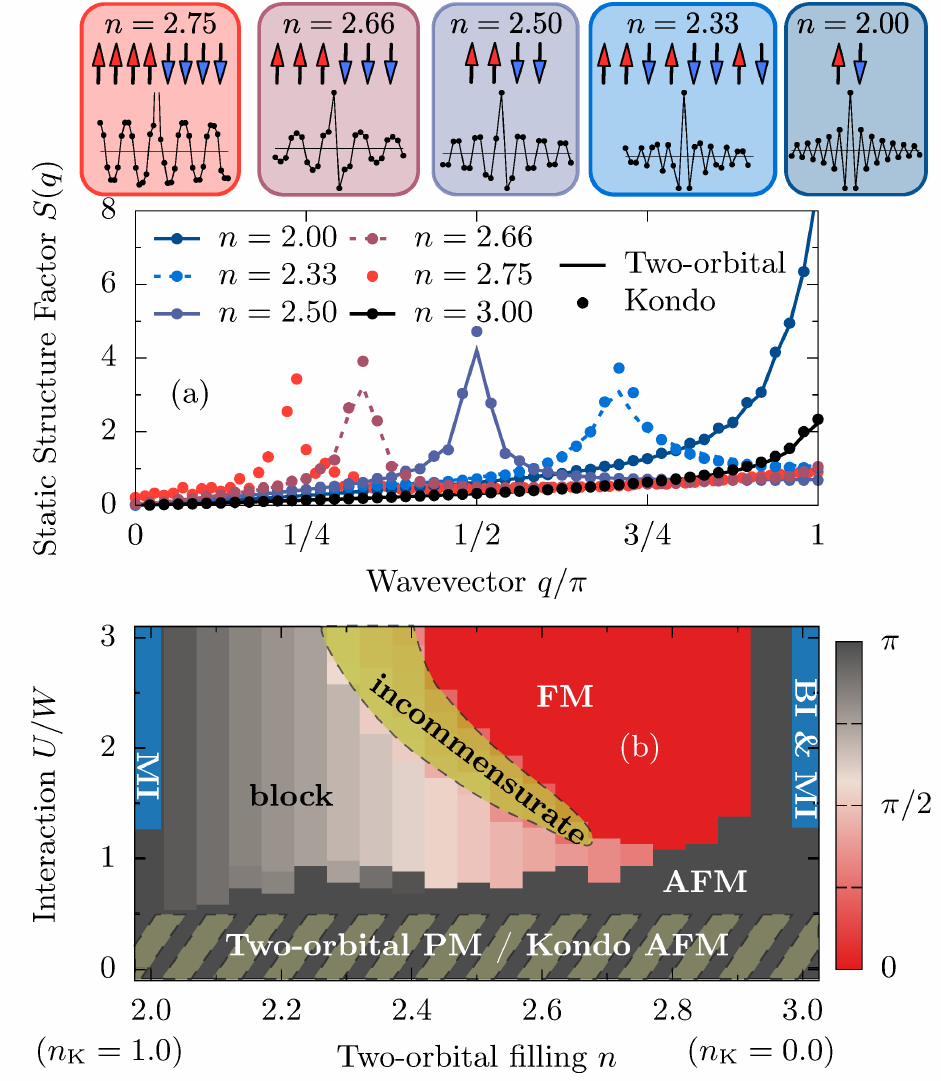}
\caption{(Color online) (a) Static structure factor $S(q)$ for various fillings $n$ at fixed $U=W$, for the two-orbital Hubbard (lines) and generalized Kondo-Heisenberg (points) models, both using $L=48$ sites ($L=72$ for $n=2.75$, see text for details). Real-space correlation functions $\langle \mathbf{S}_{\ell} \cdot \mathbf{S}_{L/2}\rangle$ of the two-orbital Hubbard model and sketches of the spin alignment are also presented at the top. (b) Interaction vs filling magnetic phase diagram of the generalized Kondo-Heisenberg model on $L=48$ sites. The dashed-shaded area represents the region where the mapping is not valid.}
\label{sq_nh}
\end{figure}

\begin{figure}[!htb]
\includegraphics[width=1.0\columnwidth]{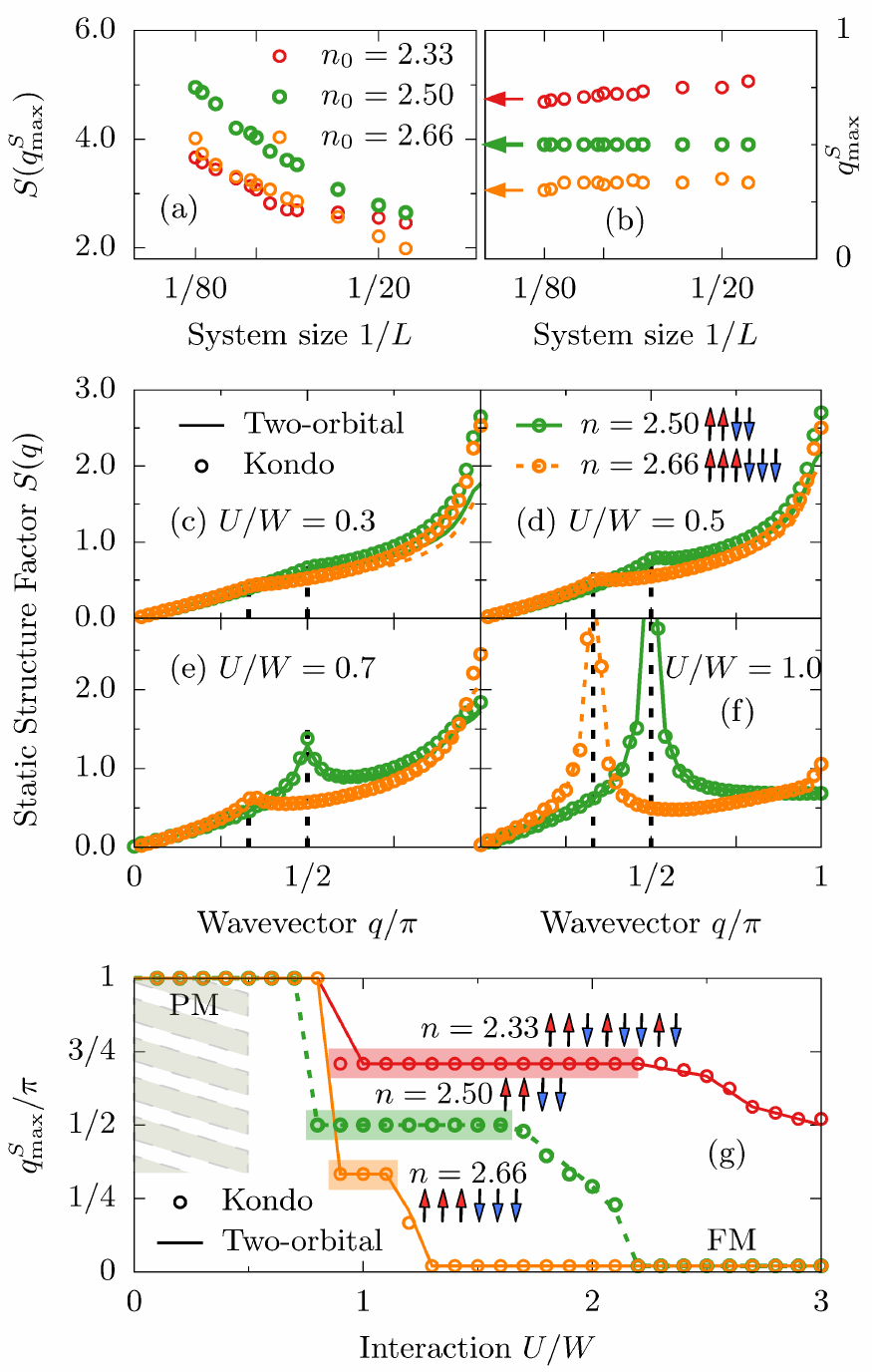}
\caption{(Color online) (a,b) Finite-size scaling of (a) the spin structure factor maximum $S(q^S_{\mathrm{max}})$ and (b) position of $q^S_{\mathrm{max}}$, calculated for $U/W=1$ using the generalized Kondo-Heisenberg model. Arrows in (b) represent the $L\to \infty$ limit $\pi n_0$. (c-f) $U$ dependence of $S(q)$ for $n=2.50$ and $2.66$. Lines (circles) represent results using the two-orbital Hubbard model (generalized Kondo-Heisenberg model) and $L=48$ sites. Dashed horizontal lines represent $2k_{\mathrm{F}}=\pi n$. (g) $U$ dependence of $q^S_{\mathrm{max}}$ for the two-orbital Hubbard model (lines) and generalized Kondo-Heisenberg model (circles) at fillings $n=2.33\,,2.50$, and $2.66$. The dashed-shaded area represents the region where the mapping is not valid.}
\label{phase_kondo}
\end{figure}

{\it Block magnetism.} Previous efforts \cite{Rincon2014,Herbrych2018} showed that the OSMP can display exotic magnetic properties, such as AFM coupled FM islands (the spin-block phase). Here, one of our main results is that the magnetic pattern of the block-OSMP is {\it not} limited to the $\pi/2$-phase, but remarkably the OSMP can support a variety of spin patterns previously unknown. Figure~\ref{sq_nh}(a) illustrates the filling $n$ dependence of the total spin structure factor $S(q)$ \cite{comment4}. Several conclusions can be obtained from the displayed results at $U=W$: (i) below half-filling, $n<2$, the (weak) maximum of $S(q)$ at $q=\pi$ indicates a paramagnetic state with short-range spin staggered tendencies. (ii) Entering the OSMP phase, $2<n<3$, robust correlations develop with well-defined peaks in $S(q)$ at $q^S_{\mathrm{max}}$ [see Fig.~\ref{phase_kondo}(a,b) and \cite{supp} for finite-size scaling]. In this region $q^S_{\mathrm{max}}$ strongly depends on $n$, and decreases as $n$ increases [see also Fig.~\ref{ocu_nh}(f)]. (iii) For $n>3$ paramagnetism is recovered with a (weak) maximum at $q=\pi$ \cite{comment1}. Although the strong spin correlations unveiled here resemble long-range order, it is expected that in 1D eventually they would decay slowly as a power law. However, weak couplings perpendicular to the chains/ladders should stabilize the unveiled orders into long-range patterns with the same local block order as reported here.

Consider now the real-space correlation functions \mbox{$\langle\mathbf{S}_\ell \cdot \mathbf{S}_{L/2}\rangle$} [Fig.~\ref{sq_nh}(a), top]. Starting with $n=2$, the structure factor has a ``standard'' $\pi$-AFM staggered spin pattern, common of Mott insulators. However, at $n\simeq2.3$ an unexpected novel spin pattern emerges involving $1$- and $2$-sites FM islands of the form $\uparrow\uparrow\downarrow\uparrow\downarrow\downarrow\uparrow\downarrow$. At \mbox{$n=2.5$} the previously known AFM coupled FM blocks of two spins, $\uparrow\uparrow\downarrow\downarrow$, is stabilized. Increasing further the electron doping, the magnetic islands continue growing in size. On the considered $L=48$ lattice, the largest new pattern observed contains FM blocks with three spins ($\uparrow\uparrow\uparrow\downarrow\downarrow\downarrow$) for $n\simeq2.66$. Increasing $n$ further, $S(q)$ suddenly reaches a maximum again at $q^S_{\mathrm{max}}=\pi$. As shown later, for large enough $L$, larger FM islands were observed \cite{comment2}, albeit in narrower regions of the parameter space.

\begin{figure}[!htb]
\includegraphics[width=1.0\columnwidth]{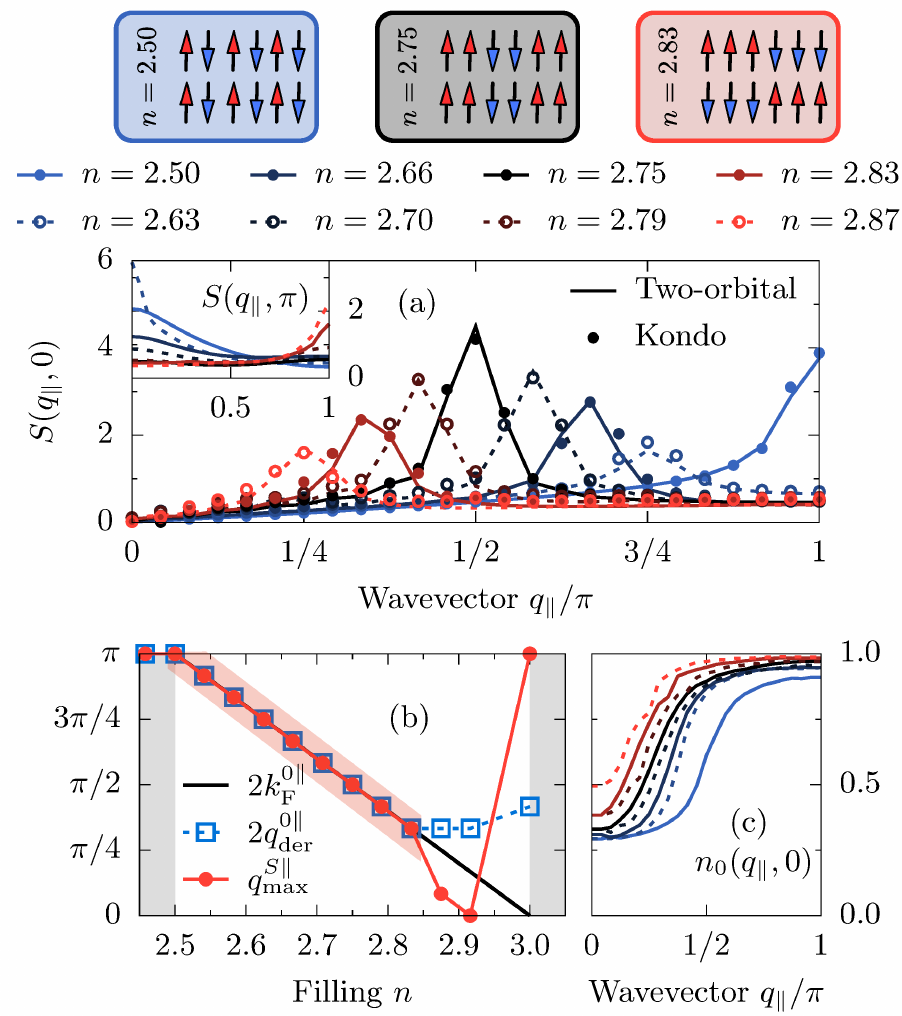}
\caption{(Color online) (a) $n$ dependence of $S(q_\parallel,0)$ (bonding component) at \mbox{$U=W_{\mathrm{L}}$} using the two-orbital Hubbard ladder (lines) and the generalized Kondo-Heisenberg (points) models, both on $L=48$ sites ($24$ rungs). The inset depicts the antibonding component $S(q_\parallel,\pi)$. Schematics of the spin configurations are also presented. (b) Filling $n$ dependence of $q^{S\parallel}_{\mathrm{max}}$, \mbox{$2k^{0\parallel}_{\mathrm{F}}=2\pi n_0$}, and $2q^{0\parallel}_{\mathrm{der}}$. Block-OSMP appears as a colored area. (c) Momentum distribution function $n_0(q_\parallel,0)$ of the $\gamma=0$ orbital in the bonding $q_\perp=0$ sector. The legend is the same as in (a).}
\label{ladder_nh}
\end{figure}

{\it Effective model.} To better understand these surprisingly rich new magnetic structures, let us consider an effective Hamiltonian. Because in the OSMP the double occupancy of the localized orbital $\gamma=1$ with $n_1=1$ is highly suppressed, it is natural to restrict - using the Schrieffer-Wolff transformation \cite{Schrieffer1966} - the Hilbert space of Eq.~$\eqref{hamhub}$ to the subspace with strictly one electron per site at $\gamma=1$ orbital. The formal derivation is in \cite{supp}, and here we just present the final result, i.e., the generalized Kondo-Heisenberg model defined as,
\begin{eqnarray}
H_{\mathrm{K}}&=&t_{\mathrm{00}}\sum_{\ell,\sigma}
\left(c^{\dagger}_{0,\ell,\sigma}c^{\phantom{\dagger}}_{0,\ell+1,\sigma}+\mathrm{H.c.}\right)
+U\sum_{\ell}n_{0,\ell,\uparrow}n_{0,\ell,\downarrow}\nonumber\\
&+&K\sum_{\ell}\mathbf{S}_{1,\ell} \cdot \mathbf{S}_{1,\ell+1}
-2J_{\mathrm{H}}\sum_{\ell}\mathbf{S}_{0,\ell} \cdot \mathbf{S}_{1,\ell}\,,
\label{hamkon}
\end{eqnarray}
with $K=4t_{11}^2/U$. The electronic filling of the effective Hamiltonian is either $n_{\mathrm{K}}=n-1$ or $n_{\mathrm{K}}=3-n$ due to the particle-hole symmetry of Eq.~\eqref{hamkon}. We tested the accuracy of Eq.~\eqref{hamkon} by comparing results for $S(q)$ in a wide range of parameters. From Figs.~\ref{sq_nh}(a), ~\ref{phase_kondo}, and ~\ref{ladder_nh}(a), clearly the magnetic properties of the multi-orbital Hamiltonian in the block-OSMP are accurately reproduced by the generalized Kondo-Heisenberg model \cite{comment5}. Furthermore, due to the small Hilbert space of the latter, we can accurately study much larger systems and stabilize larger blocks. In Fig.~\ref{sq_nh}(a) we show results for $L=72$, $U/W=1.02$, $J_{\mathrm{H}}/U=0.27$, and $n_{\mathrm{K}}=0.25$ which exhibit the block of 4-spins \cite{comment6}.

For $U=0$ (and $K=0$) the effective Hamiltonian resembles the widely-studied Kondo-Heisenberg (Kondo) model. In this framework, we can understand intuitively the origin of the magnetic blocks \cite{Batista1998,Garcia2000,Biermann2005}. At half-filling $n_{\mathrm{K}}=1$ and for a strong Kondo (Hund) coupling, $J_{\mathrm{H}}\gg K\,, t_{00}$, the $\gamma=0$ electrons form local Kondo interorbital triplets with localized spins. Increasing the electrons' mobility, $t_{00}$, leads to the double-exchange FM ordering \cite{Dagotto2001}. On the other hand, when $t_{00}\gg J_{\mathrm{H}}\,,K$ one can observe (short-range) features in the spin spectrum upon doping at twice the Fermi vector ($2k_{\mathrm{F}}$) of the itinerant orbitals, Fig.~\ref{phase_kondo}(c-d). In the OSMP, when $J_{\mathrm{H}}\sim K\sim{\cal O}(t_{00})$, the $2k_{\mathrm{F}}$ instability of itinerant orbitals is amplified leading to spin (quasi-)long-range order with maximum at $S(2k_{\mathrm{F}})$ [Fig.~\ref{phase_kondo}(e-f)]. As a consequence, competition of double- and super-exchange mechanisms leads to the formation of block magnetic islands. We tested the above ``$2k_{\mathrm{F}}$ prediction'' of the maximum of the spin structure factor by comparing $q^S_{\mathrm{max}}$ with the $U\to0$ limit in 1D, i.e., $2k^0_{\mathrm{F}}=\pi n_{0}$. Also, we calculated the Fermi vector position directly from the momentum distribution function $n_{\gamma}(q)$ via the maximum $q^0_{\mathrm{der}}$ of $\mathrm{d}n_{0}(q)/\mathrm{d}q$ [Fig.~\ref{ocu_nh}(e) and \cite{supp}]. It is evident from Fig.~\ref{ocu_nh}(c) that the block-magnetism follows the Fermi vector of the $\gamma=0$ orbital \cite{comment3}. It is important to remark that the above ``$2k_{\mathrm{F}}$-stabilization'' is an {\it emergent phenomena} unique of multi-orbital systems. Doping the single-band Hubbard model leads only to short-ranged ordered spin correlations, retaining their $\pi$-AFM character. This is strikingly different from the behavior reported here in multi-orbital models because the Hund coupling between subsystems induces novel magnetic block-states. 

As discussed above, the effective Hamiltonian Eq.~\eqref{hamkon} accurately describes the OSMP magnetic phases, and is in good agreement with the rich island-physics of Kondo-lattice Hamiltonians unveiled before \cite{Batista1998,Aliaga2001,Xavier2002,Garcia2002,Garcia2004}. This allows us to create a detailed magnetic phase diagram of the effective model. In Fig.~\ref{sq_nh}(b) we present the $n$--$U$ dependence of $q^S_{\mathrm{max}}$ using $L=48$ sites: (i) at $U<W/2$, where the mapping should not work, the system is paramagnetic; (ii) at $n<2.5$ for all considered $U$'s and for $U\simeq W$ at $n\lesssim 2.7$ we found the novel stable blocks of various sizes, depending on $k_{\mathrm{F}}\propto n$; (iii) finally at $U\gg W$ the system is ferromagnetic for all $2<n<3$ [see also Fig.~\ref{phase_kondo}(g)]. Furthermore, in between the FM and block phases we observed a narrow incommensurate magnetic region which will be studied in the future.

{\it Ladder geometry.} To confirm the robustness of our findings and to bring our results closer to experimental compounds, such as AFe$_2$X$_3$ \cite{Caron2011,Nambu2012,Caron2012,Dong2014,Mourigal2015,Hawai2015,Chi2016,Wang2017}, consider now two-orbital ladder systems. The kinetic part of the Hamiltonian is defined with isotropic hoppings \mbox{$t^{\parallel}_{\gamma\gamma^\prime}=t^{\perp}_{\gamma\gamma^\prime}=t_{\gamma\gamma^\prime}$} and $\Delta=1.6\,\mathrm{eV}$ (kinetic energy bandwidth $W_{\mathrm{L}}=3.55\,\mathrm{eV}$), while the remaining interactions are as in Eq.~\eqref{hamhub}. Figure~\ref{ladder_nh}(a) presents the bonding $q_{\perp}=0$ component of the spin structure factor $S(q_\parallel,0)$ for $U=W_{\mathrm{L}}$. Consistent with the 1D predictions, the maximum of $S(q_\parallel,0)$ strongly depends on filling. At $n=2.5$ the system is in a $\pi$-AFM state, and with increasing filling spin blocks start to develop. Interestingly, our results indicate that as $n\to 2.5$ the legs are FM aligned. However, as $n\to3$ a novel AFM ordering between the legs develop, while the FM islands involve three spins \cite{danilo2006}. The latter may arise from competing double-exchange-FM vs AFM tendencies coming from the localized spins and Fermi instability $k_{\mathrm{F}}$, namely the energy of the large FM blocks as $n\to3$ is reduced by the rung AFM arrangement. Thus, we speculate that the ladder geometry can stabilize even larger magnetic blocks due to the AFM ordering between legs.

Consider now the filling dependence of the maximum of the ladder spin structure factor $q^{S\parallel}_{\mathrm{max}}$. As in 1D, $q^{S\parallel}_{\mathrm{max}}$ follows the Fermi vector $q^{0\parallel}_{\mathrm{der}}$ estimated from the momentum distribution function [Fig.~\ref{ladder_nh}(b)]. Although the region where the block magnetism is observed, $2.5<n<3$, is smaller compared with chains, this can be explained considering the ladder $U\to 0$ limit prediction of Fermi vectors. In this case, for an isotropic ladder ($t^{\parallel}_{00}=t^{\perp}_{00}$) and $n>1.5+1$, the itinerant orbital Fermi vector is $2k_{\mathrm{F}}^{0\parallel}=2\pi n_0$ (as opposed to $2k_{\mathrm{F}}^{0}=\pi n_0$ for a chain). Figures~\ref{ladder_nh}(b,c) indicate excellent agreement between $2k_{\mathrm{F}}^{0\parallel}$ and the numerically evaluated $2q^{0\parallel}_{\mathrm{der}}$ [$\mathrm{d}n(q_\parallel,0)/\mathrm{d}q$ maximum] in the OSMP. Furthermore, as in chains, the block-magnetism follows the $2k_{\mathrm{F}}^{0\parallel}$ prediction.

{\it Conclusions.} We have shown that the multi-orbital Hubbard model in the OSMP regime supports spin-block magnetism of various novel sizes and shapes, depending on filling and lattice geometry. Moreover, we also derived an effective OSMP Hamiltonian, the generalized Kondo-Heisenberg model, which describes all magnetic phases accurately. The observed spin structures are related to the $2k_{\mathrm{F}}$ of the metallic electron bands, but spin blocks are much more sharply defined than they would be in the mere sinusoidal structure arising from a weak-coupling Fermi-surface instability [see real-space spin correlations in Fig.~\ref{sq_nh}(a) top] . We believe that the strongly correlated nature of the localized spins, due to its narrow bandwidth \cite{Li2016}, enhances the $2k_{\mathrm{F}}$ instability in a manner only possible in OSMP regimes. Our predictions could be relevant within the 123 families of iron-based materials and can be confirmed by INS experiments. But we remark that our results are generic and could apply to any quasi-1D quantum material in an OSMP regime.

\let\oldaddcontentsline\addcontentsline
\renewcommand{\addcontentsline}[3]{}
\begin{acknowledgments}
We thank C.~Batista and N.~Kaushal for fruitful discussions. J.~Herbych, N.~D.~Patel, A.~Moreo, and E.~Dagotto were supported by the US Department of Energy (DOE), Office of Science, Basic Energy Sciences (BES), Materials Sciences and Engineering Division. J.~Herbrych acknowledges also partial support by the Polish National Agency of Academic Exchange (NAWA) under contract PPN/PPO/2018/1/00035. G.~Alvarez was partially supported by the Center for Nanophase Materials Sciences, which is a DOE Office of Science User Facility, and by the Scientific Discovery through Advanced Computing (SciDAC) program funded by U.S. DOE, Office of Science, Advanced Scientific Computing Research and Basic Energy Sciences, Division of Materials Sciences and Engineering. J.~Heverhagen and M.~Daghofer were supported by the Deutsche Forschungsgemeinschaft, via the Emmy-Noether program (DA 1235/1-1) and FOR1807 (DA 1235/5-1) and by the state of Baden-W\"{u}rttemberg through bwHPC.
\end{acknowledgments}

\let\addcontentsline\oldaddcontentsline
\clearpage
\appendix
\setcounter{page}{1}
\setcounter{figure}{0}
\setcounter{equation}{0}
\newcommand{\rom}[1]{\uppercase\expandafter{\romannumeral #1\relax}}
\renewcommand{\citenumfont}[1]{S#1}
\renewcommand{\bibnumfmt}[1]{[S#1]}
\renewcommand{\thepage}{S\arabic{page}}
\renewcommand{\thefigure}{S\arabic{figure}}
\renewcommand{\theequation}{S\arabic{equation}}
\onecolumngrid
\begin{center}
{\bf \uppercase{Supplementary Information}} for:\\
\vspace{10pt}
{\bf \large Novel Magnetic Block States in Low-Dimensional Iron-Based Superconductors}\\
\vspace{10pt}
by J. Herbrych, J. Heverhagen, N. Patel, G. Alvarez, M. Daghofer, A. Moreo, and E. Dagotto
\end{center}
\tableofcontents
\vspace{40pt}
\twocolumngrid

\section{S1. Analysis of the OSMP}

\subsection{Charge fluctuations}

\begin{figure}[!htb]
\includegraphics[width=1.0\columnwidth]{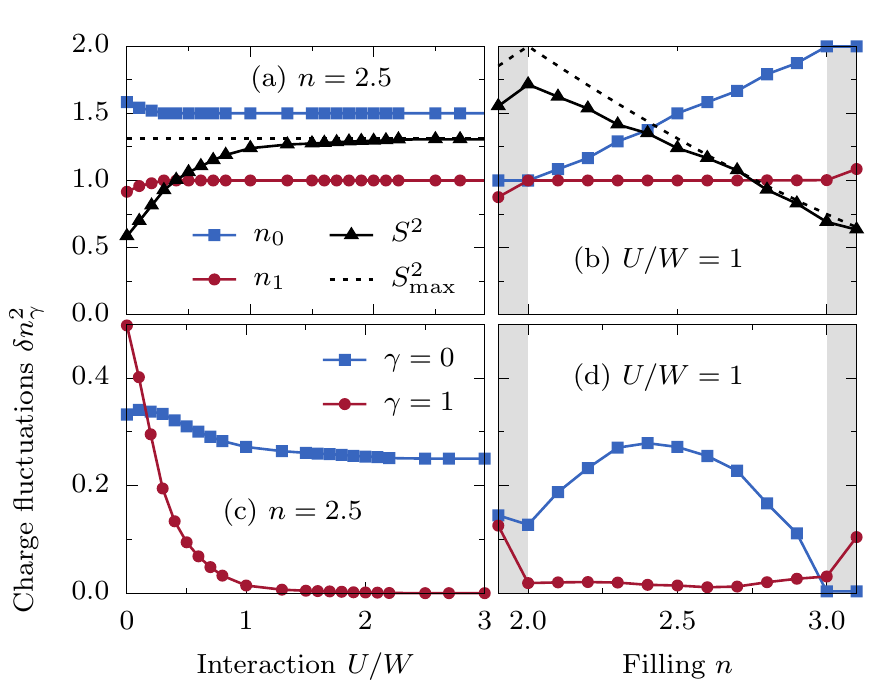}
\caption{(Color online) (a,c) Interaction $U$ and (b,d) filling $n$ dependence of the total magnetic moment per site $\langle S_\ell^2 \rangle$, orbital-resolved occupation numbers $n_{\gamma}$, and charge fluctuations $\delta n^2_{\gamma}$ of the two-orbital Hubbard model Eq.~1 of the main text, calculated using $L=48$ sites. In (a,c) $n=2.5$ and in (b,d) $U=W$. The black dashed lines represent the largest possible magnetic moment $S^2_{\mathrm{max}}$.}
\label{ocu_nh_add}
\end{figure}

Figure~\ref{ocu_nh_add} shows the interaction $U$ and filling $n$ dependence of the orbital $\gamma$ resolved charge fluctuations
\begin{equation}
\delta n^2_\gamma=\langle n^2_\gamma\rangle-\langle n_\gamma\rangle^2\,,
\end{equation}
together with other single-site expectation values presented already in the main text. Consistent with the OSMP, we find vanishing $\delta n^2_1$ when $n_1=1$. Simultaneously, $\delta n^2_0\ne0$ indicating the itinerant (metallic) character of the $\gamma=0$ orbital.

\subsection{Density of states}

\begin{figure}[!htb]
\includegraphics[width=1.1\columnwidth]{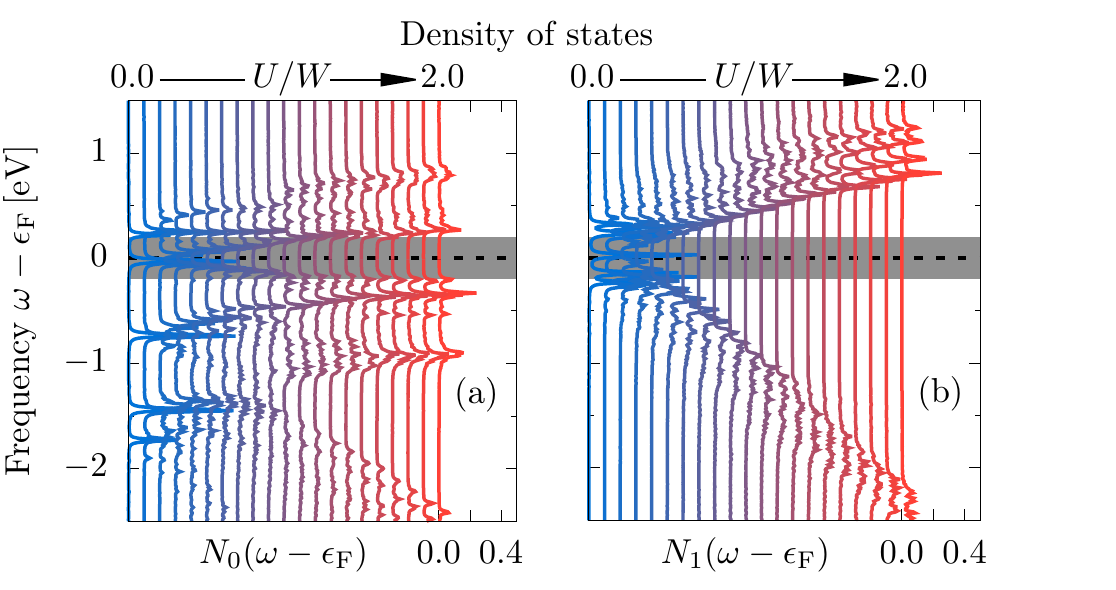}
\caption{(Color online) Interaction dependence of the orbital-resolved density of states $N_\gamma(\omega)$. Results for (a) $\gamma=0$ and (b) $\gamma=1$ calculated for filling $n=2.5$ and system size $L=8$ using Lanczos diagonalization (periodic boundary conditions were used). The shaded areas represent $\pi/L$ estimates of finite-size effects.}
\label{Sak_nh}
\end{figure}

To strengthen our analysis of the orbital selective Mott phase (OSMP), in Fig.~\ref{Sak_nh} we present the orbital-resolved density of states (DOS) for filling $n=2.5$ calculated as
\begin{equation}
N_\gamma(\omega)=-\frac{1}{\pi}\sum_{k}\,\mathrm{Im}\,G_{\gamma}(k,\omega),
\end{equation}
where $G(k,\omega)$ is a single-particle Green's function of the $\gamma$ orbital electrons. 

Let us start our analysis of the DOS with the $\gamma=1$ orbital, Fig.~\ref{Sak_nh}(b). Consistent with Mott insulator behavior, upon increasing the interaction $U$ one can observe two Hubbard bands already at $U_{\mathrm{c}}\simeq W/2$, with no states at the Fermi level $\epsilon_{\mathrm{F}}$. For this value of the interaction note that the $\gamma=1$ orbital is already singly-occupied and that the effective interaction (corresponding to a single-band picture) is $U_{\mathrm{c}}/t_{11}\simeq7$. As a consequence, the $\gamma=1$ orbital behaves like a Mott insulator. On the other hand, the $\gamma=0$ orbital has a nonzero Fermi-level DOS, $N(\epsilon_{\mathrm{F}})\ne0$ (at $\epsilon_{\mathrm{F}}$, or, due to small finite-size effects, in the vicinity of $\epsilon_{\mathrm{F}}$), for all presented values of $U$, see Fig.~\ref{Sak_nh}(a). Such a behavior is consistent with itinerant electrons on orbital $\gamma=0$ and with the metallic character of this orbital.

\section{S2. Additional results for the spin structure factor}

\subsubsection{Finite-size scaling}

\begin{figure}[!htb]
\includegraphics[width=1.0\columnwidth]{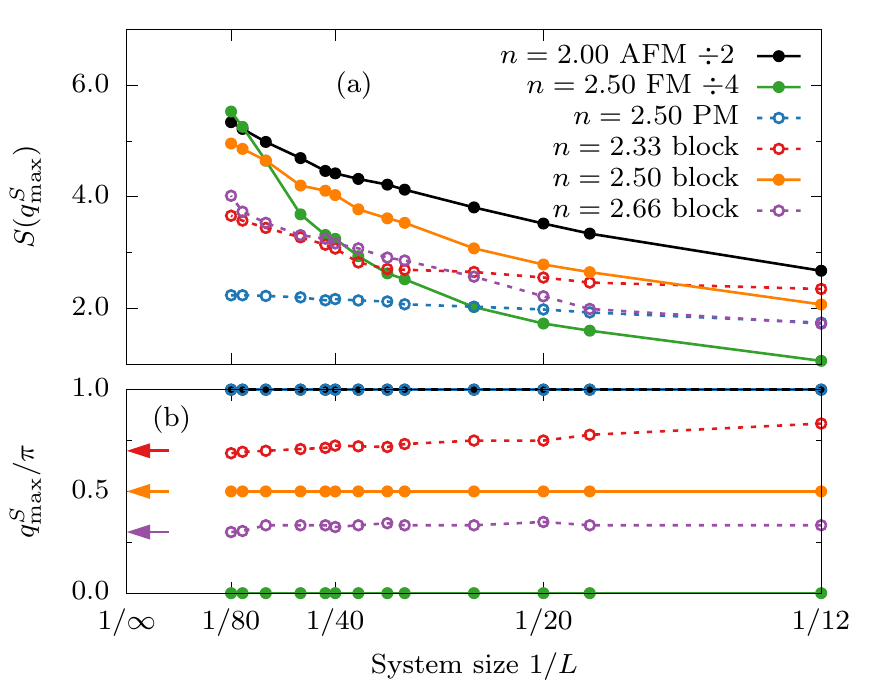}
\caption{(Color online) Finite-size scaling of (a) the maximum of the spin structure factor $S(q^S_{\mathrm{max}})$ and (b) position of $q^S_{\mathrm{max}}$. The presented results are: $\pi$-AFM for $n=2\,,U/W=1$; magnetic-block states for $n=2.33\,,2.50\,,2.66$ and $U/W=1$; FM-state for $n=2.5\,,U/W=4$; paramagnetic (PM) state for $n=2.5\,,U/W=0.5$. Arrows in panel (b) represent the $L\to\infty$ limit given by $\pi n_0$.}
\label{Ssize_nh}
\end{figure}

Figure~\ref{Ssize_nh}(a) presents the finite-size scaling of the maximum of the spin structure factor $S(q^S_{\mathrm{max}})$ for various doping and interaction $U$ values. Consistent with the discussion presented in the main text, all block-magnetic phases of OSMP manifest a robust magnetic order, i.e., $U=W$ block-islands for $n=2.33\,,2.50$, and $2.66$. For consistency, we present also results for $\pi$-AFM order ($U/W=1\,,n=2$), FM-order ($U/W=4\,,n=2.5$), and paramagnetic (PM) short-range order for $U/W=0.5\,,n=2.5$.

Furthermore, Fig.~\ref{Ssize_nh}(b) shows the finite-size scaling of the position of the maximum $q^S_{\mathrm{max}}$. It is evident from the presented results that our block states are not an artifact of finite-size systems: there are well-defined values in the $L\to\infty$ limit given by $\pi n_0$ (arrows).

\subsubsection{Results with finite interorbital hybridization}

\begin{figure}[!htb]
\includegraphics[width=1.0\columnwidth]{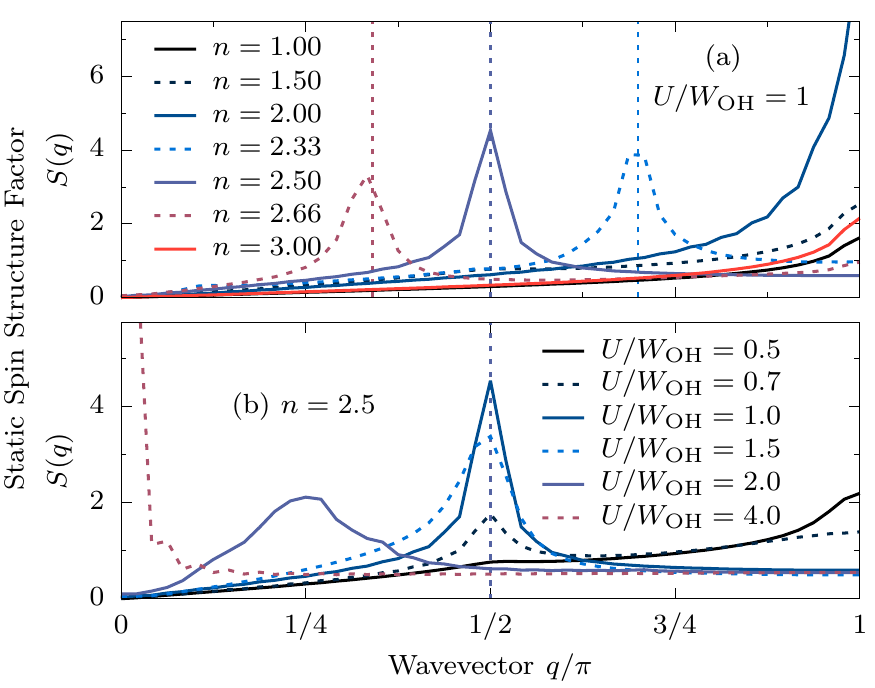}
\caption{(Color online) Total static structure factor $S(q)$ for (a) various fillings $n$ and fixed values of interaction $U=W_{\mathrm{OH}}$ and (b) for various strengths of interaction $U$ and fixed filling $n=2.5$, as calculated using $L=48$ sites and a two-orbital Hubbard model with a finite value of the interorbital hybridization $t_{01}=0.1$. Dashed vertical lines indicate the value of $2k^0_{\mathrm{F}}=\pi n_0$.}
\label{Ssq_wh}
\end{figure}

Figure~\ref{Ssq_wh} presents results obtained for the two-orbital Hubbard model described in the main text with a finite interorbital hybridization $t_{01}=0.1$ (the kinetic energy bandwidth is now $W_{\mathrm{OH}}=2.28$). It is clear that a finite $t_{01}$ does not change the main features of the spin structure factor $S(q)$ (as compared with Fig.~2(a) and Fig.~3(a) of the main text).

\subsubsection{Results for three-orbital Hubbard model}

\begin{figure}[!htb]
\includegraphics[width=1.0\columnwidth]{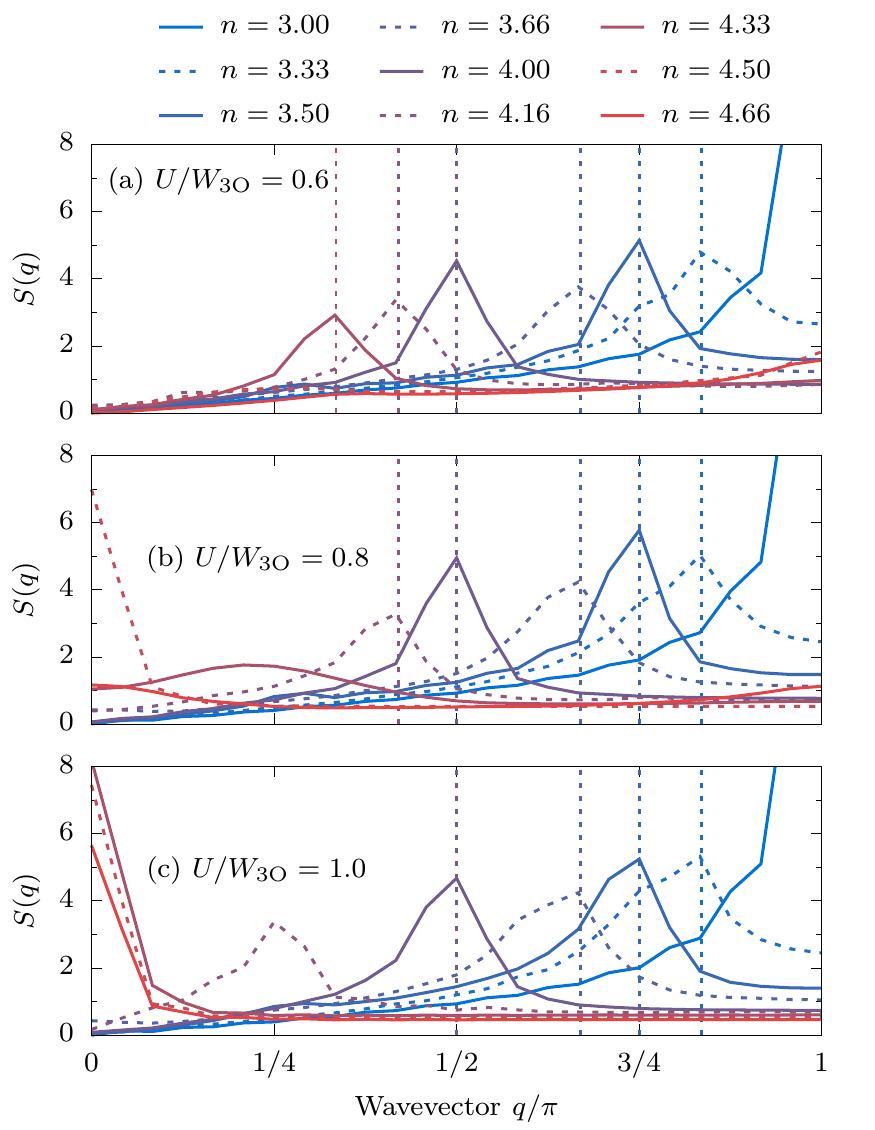}
\caption{(Color online) Total static structure factor $S(q)$ for various fillings $n$ for the three-orbital 
Hubbard model as calculated using $L=24$ sites and interaction (a) $U/W_{\mathrm{OH}}=0.6$, (b) $U/W_{\mathrm{OH}}=0.8$, and (c) $U/W_{\mathrm{OH}}=1.0$. Dashed vertical lines indicate the value of \mbox{$2\overline{k}_{\mathrm{F}}=\pi (n_0+n_1)/2$}.}
\label{Ssq_3o}
\end{figure}

Figure~\ref{Ssq_3o} displays results for $S(q)$ as calculated for the three-orbital Hubbard model. The used parameters are: \mbox{$t_{00}=t_{11}=-0.5$}, $t_{22}=-0.15$, $t_{01}=t_{02}=0.1$, $\Delta_0=-0.1$, $\Delta_1=0$, and $\Delta_0=0.8$, all in eV units. The kinetic energy bandwidth is $W_{\mathrm{3O}}=2.45\,\mathrm{eV}$. Panels (a), (b), and (c) of Fig.~\ref{Ssq_3o} depict results for $U/W_{\mathrm{3O}}=0.6\,,0.8$, and $1.0$, respectively.

In the three-orbital system, $\gamma=2$ is localized while orbitals $\gamma=0$ and $\gamma=1$ are itinerant. As a consequence, the Fermi vector related to the maximum of the spin structure factor can be calculated as $2\overline{k}_{\mathrm{F}}=\pi (n_0+n_1)/2$.

\section{S3. Mapping to the generalized Kondo-Heisenberg model}

Our aim in this section is to find a unitary operator ${\cal S}$ via a Schrieffer-Wolff transformation $H_{{\cal S}}=\mathrm{e}^{i {\cal S}}H_{\mathrm{H}}\mathrm{e}^{-i {\cal S}}$ such that, at a given order, eliminates transitions between states with a different number of doubly occupied sites for the localized orbital $\gamma=1$. Next, because in the OSMP the states with empty and double occupied sites of the $\gamma=1$ orbital are highly suppressed [see Fig.~1(e,f) of the main text], we can restrict the Hilbert space of $H_{{\cal S}}$ to the subspace with (strictly) one electron per site at this orbital. In other words, we search for the Hamiltonian 
\begin{equation}
H_{{\cal S}}=H_{\mathrm{H}}+ i[{\cal S},H_{\mathrm{H}}]+{\cal O}({\cal S}^2)
\label{Ssw}
\end{equation}
for which $N_1=\sum n_{1,\ell}$ is a good quantum number up to second order in ${\cal S}$, $[H_{{\cal S}},N_{1}]={\cal O}({\cal S}^2)$, and $N_1^{\mathrm{d}}=\sum n^{d}_{1,\ell}={\cal O}({\cal S}^2)$ with $n^{d}_{\gamma,\ell}$ as a doublon number operator.

Let us start by rewriting the two-orbital Hubbard model [Eq.~(1) of the main text] in the following form: $H_{\mathrm{H}}=H_{0}+H_{1}+H_{\mathrm{mix}}$. Here, $H_{\gamma}$ is the portion of the Hamiltonian which contains the terms solely related to the orbital $\gamma$ (terms similar to single-band Hubbard Hamiltonian). $H_{\mathrm{mix}}$ contains the terms which mix the orbitals, i.e., the interorbital interaction \mbox{$U^\prime=U-5J_{\mathrm{H}}/2$}, Hund coupling, pair-hopping, and the kinetic term containing the orbital hybridization $t_{01}$. However, only the last two terms can change the doublon number in $\gamma=1$. Because (i) we neglected the hybridization term, $t_{01}=0$, and (ii) in the OSMP phase the double occupancy in the $\gamma=1$ orbital is suppressed and the pair-hopping have negligible contribution, we can assume $[H_{\mathrm{mix}},N_1]\simeq0$.

In the rest, consider $H_{1}$ as
\begin{eqnarray}
H_{1}=t_{11}H_{1}^{0}+t_{11}H_{1}^{+}+t_{11}H_{1}^{-}+UH_{1}^{\mathrm{U}}\,,
\end{eqnarray}
where $H_{1}^{0}$ contains the terms that do not change the double occupancy in the $\gamma=1$ orbital (hopping of the holes and doublons), $H_{1}^{+}$ ($H_{1}^{-}$) contains the terms which increase (decrease) the doublon number on the $\gamma=1$ orbital, and $H_{1}^{\mathrm{U}}$ which is the Hubbard-$U$ term on the $\gamma=1$ orbital. It can be shown that
\begin{eqnarray}
H_{1}^{+}&=&-\sum_{\ell,\sigma}
\left[n_{1,\ell,-\sigma}c^{\dagger}_{1,\ell,\sigma}c^{\phantom{\dagger}}_{1,\ell+1,\sigma}\widetilde{n}_{1,\ell+1,-\sigma}+\mathrm{H.c.}\right]\,,\nonumber\\
H_{1}^{0}&=&-\sum_{\ell,\sigma}\left[
n_{1,\ell,-\sigma}c^{\dagger}_{1,\ell,\sigma}c^{\phantom{\dagger}}_{1,\ell+1,\sigma}n_{1,\ell+1,-\sigma}\right.\nonumber\\
&&\left.+
\widetilde{n}_{1,\ell,-\sigma}c^{\dagger}_{1,\ell,\sigma}c^{\phantom{\dagger}}_{1,\ell+1,\sigma}\widetilde{n}_{1,\ell+1,-\sigma}+\mathrm{H.c.}\right]\,,
\end{eqnarray}
where $\widetilde{n}_{1,\ell,\sigma}=(1-n_{1,\ell,\sigma})$ and also 
$H_{1}^{-}=(H_{1}^{+})^\dagger$. Collecting the terms in Eq.~(\ref{Ssw}) we obtain
\begin{eqnarray}
H_{{\cal S}}&=&
H_{0}+H_{\mathrm{mix}}+t_{11}H_{1}^{0}+t_{11}H_{1}^{+}+t_{11}H_{1}^{-}+UH_{1}^{U}\nonumber\\
&+& i[{\cal S},H_{\mathrm{H}}]+{\cal O}({\cal S}^2)\,.
\label{Shef}
\end{eqnarray}
In order to eliminate the transitions between the states with different doublon number on the $\gamma=1$ orbital we have to require the $ i[{\cal S},H_{\mathrm{H}}]$ term to eliminate the $H_{1}^{\pm}$ contribution. The latter can be achieved via
\begin{equation}
{\cal S}=-i\frac{t_{11}}{U}\left(H_{1}^{+}-H_{1}^{-}\right)\,,
\end{equation}
and as a consequence ${\cal O}({\cal S}^2)\propto t_{11}^3/U^2$. For the system parameters considered in Eq.(\ref{hamhub}) $t_{11}^3/U^2\ll W$ in the OSMP.

In the last step we have to evaluate $[{\cal S},H_{\mathrm{H}}]$. Some observations are in order: 
\begin{itemize}
\item[(i)] $[{\cal S},H_0]=0$, since the operator ${\cal S}$ contains only the terms involving $\gamma=1$.
\item[(ii)] It can be shown that $i U[{\cal S},H_{1}^{U}]=-(H_{1}^{+}+H_{1}^{-})$. Note that this contribution removes (as required) transitions between different doubly occupied sites at the $\gamma=1$ orbital.
\item[(iii)] The term $[{\cal S},H_{1}^{0}]$ will involve a single occurrence of the $H_{1}^{\pm}$ operator and, therefore, it will be increasing/decreasing the double occupancy at $\gamma=1$. Since the latter is not allowed in our restricted space, this term can be omitted.
\item[(iv)] At half-filling, $n_1=1$, $H_{1}^{0}$ can be also neglected since it describes the hoping of doublons and holons, not present in OSMP. 
\item[(v)] In the restricted subspace of one electron per site at the $\gamma=1$ orbital the $H_{1}^{U}$ term, together with the $U^\prime$ and pair-hopping term in $H_{\mathrm{mix}}$, can be dropped. The first two are just constants, where the last one is strictly $0$.
\item[(vi)] Also, the term proportional to $H_{1}^{+}H_{1}^{-}$ can be omitted, since it creates (at half-filling) a holon-doublon pair not allowed in the reduced Hilbert space.
\item[(vii)] $[{\cal S},H_{\mathrm{mix}}]$ will contain terms which conserves doublons ($H_{\mathrm{mix}}$) and terms which change the doublon number (${\cal S}$). As a consequence, similarly to point (iii), this contribution can be omitted in our restricted space.
\item[(viii)] The term $-t_{11}^2/UH_{1}^{-}H_{1}^{+}$ can be written as $4t_{11}^2/U\sum_{\ell} \mathbf{S}_{1,\ell} \cdot \mathbf{S}_{1,\ell+1}$ (virtual spin flips) where $\mathbf{S}_{1,\ell}$ is the total spin at site $\ell$ of orbital $\gamma=1$.
\end{itemize}
Readers will recognize that, due to the assumptions justified when addressing the OSMP regime, the above transformation is simply a ``large-$U$'' Hubbard to Heisenberg mapping involving the localized $\gamma=1$ orbital. Collecting relevant terms, and omitting the ${\cal O}(t_{11}^2/U)$ term, the above Hamiltonian can be written in the familiar form of the generalized Kondo-Heisenberg model, namely
\begin{eqnarray}
H_{\mathrm{K}}&=&t_{\mathrm{00}}\sum_{\ell,\sigma}
\left(c^{\dagger}_{0,\ell,\sigma}c^{\phantom{\dagger}}_{0,\ell+1,\sigma}+\mathrm{H.c.}\right)
+U\sum_{\ell}n_{0,\ell,\uparrow}n_{0,\ell,\downarrow}\nonumber\\
&-&2J_{\mathrm{H}}\sum_{\ell}\mathbf{S}_{0,\ell} \cdot \mathbf{S}_{1,\ell}
+K\sum_{\ell}\mathbf{S}_{1,\ell} \cdot \mathbf{S}_{1,\ell+1}\,,
\label{Shamkon}
\end{eqnarray}
with $K=4t_{11}^2/U$.

\section{S4. Additional results for the momentum distribution function}

\begin{figure}[!htb]
\includegraphics[width=1.0\columnwidth]{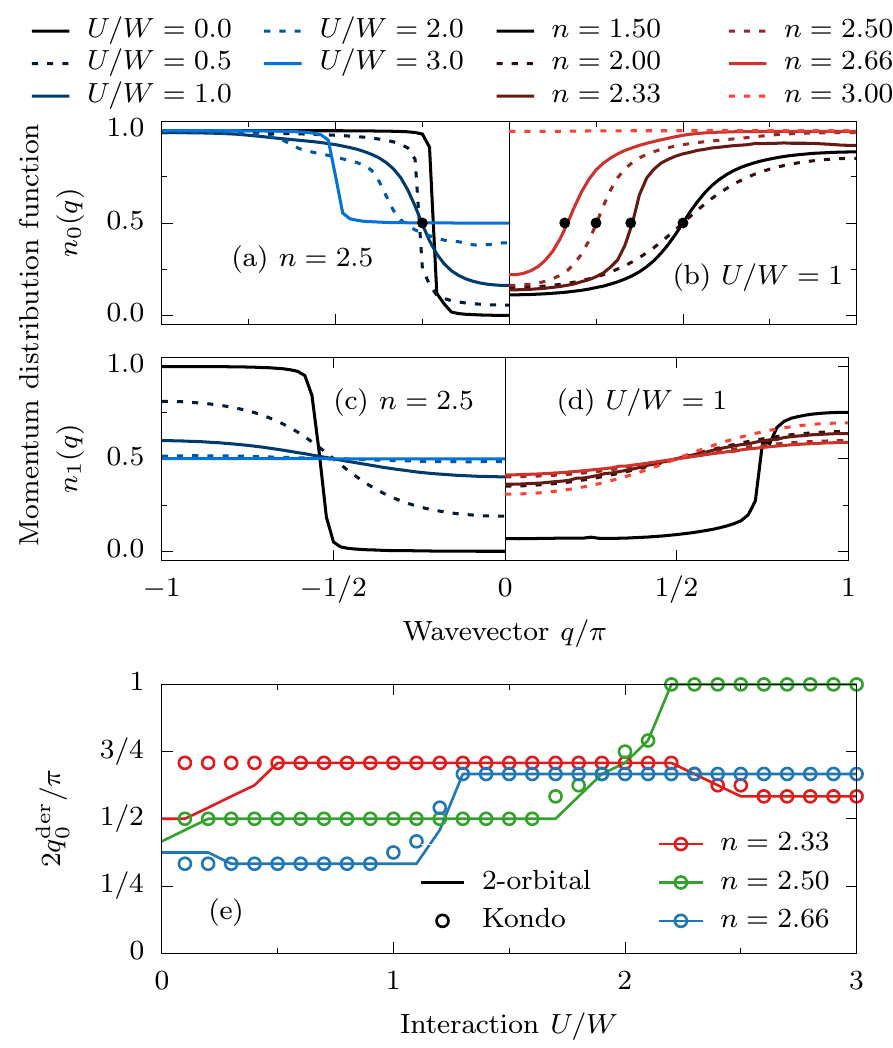}
\caption{(Color online) (a,c) Interaction $U$ and (b,d) filling $n$ dependence of the orbital-resolved momentum distribution function $n_{\gamma}(q)$ of the two-orbital Hubbard model (using $L=48$ sites). The upper (lower) row represents results for $\gamma=0$ ($\gamma=1$). Black points depict the position of the maximum of the derivative $\mathrm{d}n_{0}(q)/\mathrm{d}q$, i.e., $q^0_{\mathrm{der}}$. (e) Interaction dependence of $2q^0_{\mathrm{der}}$. Lines (points) represent results obtained with a two-orbital Hubbard model (generalized Kondo-Heisenberg model).}
\label{Snk_nh}
\end{figure}

Figure~\ref{Snk_nh} presents an analysis of the momentum distribution function
\begin{equation}
n_{\gamma}(q)=\sum_{\ell,\kappa,\sigma}\mathrm{e}^{iq(\ell-\kappa)}
\langle c^{\dagger}_{\gamma,\ell,\sigma}c^{\phantom{\dagger}}_{\gamma,\kappa,\sigma}\rangle\,.
\end{equation}
In panels (a-d), we show the interaction $U$ and filling $n$ dependence of $n_{\gamma}(q)$. Furthermore, in Fig.~\ref{Snk_nh}(e) we present the comparison between the two-orbital and generalized Kondo-Heisenberg model of $2q^0_{\mathrm{der}}$ maximum. Similarly to the case of the spin structure factor $S(q)$, we have found a perfect agreement for $U/W\gtrsim0.5$.

\begin{thebibliography}{99}
\bibitem{comment1} Note that in this doping region the orbital $\gamma=0$ is in a band-insulator state, while the electrons on the $\gamma=1$ orbital are itinerant ($\delta n^2_1>0$).
\bibitem{comment2} On the other hand, patterns with $q^S_{\mathrm{max}}\to\pi/L$ ($n\to3-1/L$) are not possible since this would lead (in the extreme) to nonphysical cases involving just one domain wall in the system (spin block of $L/2$ spins).
\bibitem{comment6} In this case we have included small Ising-like anisotropy in the localized spins, i.e., $0.83K^{\perp}=K^z$, which helps to stabilize the block pattern.
\bibitem{comment3} It is also interesting to note that $k^0_{\mathrm{F}}\ne q^0_{\mathrm{der}}$ for the same filling $n\gtrsim2.7$ for which we observe the breakdown of the magnetic blocks.
\end{thebibliography}
\end{document}